\documentclass{ws-mpla}
\usepackage[super,compress]{cite}
\usepackage{hyperref}
\hypersetup{colorlinks,urlcolor=black,citecolor=black,linkcolor=black,filecolor=black}
\usepackage{breakurl}
\usepackage[numbers,sort&compress]{natbib}
\usepackage{changepage}

\begin{document}

\thispagestyle{plain}

\title{On Scalar Cosmological Perturbations in Non-Minimally Coupled Weyl Connection Gravity}

\author{M. Lima}

\address{Departamento de Física, Instituto Superior Técnico, Universidade de Lisboa, Av. Rovisco Pais, 1049-001 Lisboa, Portugal}
\address{Centro de Análise Matemática, Geometria e Sistemas Dinâmicos, Instituto Superior Técnico, Universidade de Lisboa, Av. Rovisco Pais, 1049-001 Lisboa, Portugal}
\address{Institute Okeanos - University of the Azores, Campus da Horta, Rua Professor Doutor Frederico Machado 4, 9900-140 Horta, Portugal\\
margarida.a.lima@tecnico.ulisboa.pt}

\author{C. Gomes}

\address{Centro de Física das Universidades do Minho e do Porto, Faculdade de Ciências da Universidade do Porto, Rua do Campo Alegre s/n, 4169-007 Porto
}
\address{Institute Okeanos - University of the Azores, Campus da Horta, Rua Professor Doutor Frederico Machado 4, 9900-140 Horta, Portugal\\
claudio.gomes@fc.up.pt}

\maketitle

\begin{abstract}
We analyze a theory with non-minimal matter–curvature coupling, considering non–metricity properties with a Weyl connection. This model has the advantage of an extra force term which can mimic dark matter and dark energy, and simultaneously follow Weyl’s idea to unify gravity and electromagnetism. Indeed, Schwarzschild-like and Reissner--Nordstrøm-like black hole solutions exist in this model, leading to new features, such as an additional horizon, due to the non-metricity vector. We derive the cosmological field equations, considering a minimal coupling, and discuss preliminary results on the scalar cosmological perturbations in this model. 

\end{abstract}
\section{Introduction}

Astrophysical and cosmological data favor the existence of two dark components at large scales, namely dark matter and dark energy, which amount to around 95\% of the energy content of the Universe. This occurs provided that we maintain General Relativity (GR) as the theory of gravity. However, Einstein's model also suffers from other pathologies such as the existence of singularities, the so-called cosmological constant problem, and being non-renormalizable. Therefore, alternative theories of gravity were explored. The simplest generalization is the $f(R)$ models \cite{Sotiriou:2008rp,DeFelice:2010aj}, which can include the Starobinsky model, $f(R)=R+\alpha R^2$, which fits very well Planck's data \cite{Planck:2018inflation}. However, other extensions can be implemented \cite{Capozziello:2011et,Nojiri:2010wj,Nojiri:2017ncd}. One such realization further extends $f(R)$ gravity by including an explicit non-minimal coupling between matter and curvature \cite{Bertolami:2007}. This model generates an extra force term in the geodesics for a perfect fluid, which can mimic both dark matter in galaxies and clusters and the late-time acceleration \cite{Bertolami:2009ic,Bertolami:2011ye,Bertolami:2010cw}. It has been extensively studied in the literature \cite{Bertolami:2008perfectfluids,Olmo:2014sra,Gomes:2014,Gomes:2017,Gomes:2018gws,Azevedo:2019krx,Gomes:2020jeans,Gomes:2022quantumjeans,March:2021mqu,Bertolami:2022magnetic,Ortiz-Banos:2021jgg,BarrosoVarela:2024htf}, where it meets the stringent observational data from several sources. Moreover, we can incorporate non-metricity properties into gravity, for instance by considering the Weyl connection in a non-minimally coupled gravity model, such that spacetime is described by both the metric field and a vector field related to this property \cite{Gomes:2019weyl}. The original Weyl idea was to unify gravity and electromagnetism \cite{Weyl:1918ib}, which had problems such as the possibility of continuous and arbitrary length variation of a vector from point to point. However, the model \cite{Gomes:2019weyl} eliminates higher-order derivative terms in the metric field equations that stems from higher-order curvature terms, and has a well-behaved space-form. Moreover, it provides a viable cosmological description, even when the Weyl vector is identified as a gauge vector, and is free of Ostrogradsky instabilities \cite{Baptista:2019wdu,Baptista:2020adz}. It also admits black hole solutions that incorporate the non-metricity effects \cite{Lima:2024}, and it has a well-defined weak field limit which can coincide with the standard one from GR, in a suitable choice for the Weyl vector \cite{Gomes:2025weakfield}. Other extensions look into gravity theories whose action has a term of the form $f(R,\mathcal{A})$ \cite{Kouniatalis:2024gnr}, or resort to the so-called Weyl gravity which is built from the square of Weyl's tensor \cite{Ghilencea:2020rxc,Condeescu:2023izl,Harko:2024fnt}.

Thus, the present work aims at reviewing some results on the non-minimally coupled gravity models, and to derive both modified Friedmann equations and the cosmological perturbation equations for a minimally coupled Weyl gravity model for the first time, which is a subset of the most general non-minimally coupled Weyl connection gravity. Throughout the work, we shall use Greek letters to run over $\{0,1,2,3\}$, while roman ones run over spatial components only. Moreover, we shall use a prime to denote a time derivative in conformal time.
This work is organized as follows. In Section \ref{sec:nmc}, we present the general results from the metric non-minimally coupled model, to introduce its generalization with the Weyl connection in Section \ref{NMCWCT-section}, where we also critically discuss black hole solutions. In Section \ref{sec:modifiedFriedmann}, we derive the modified Friedmann equations in the minimal case, i.e., $f_2(\bar{R})=1$, and in Sec. \ref{sec:cosmologicalperturbations}, we resort to the scalar perturbation theory to find its consequences for this gravity model. We draw our conclusions in \ref{sec:conclusions}. 

\section{Non-Minimal Matter--Curvature Coupling Theory}\label{sec:nmc}

The action function for non-minimal coupling between curvature and matter can be formulated as \cite{Bertolami:2007}:
\begin{equation}
    S=\int \left( \kappa f_1(R)+f_2(R) \mathcal{L} \right)\sqrt{-g}~\mathrm{d}^4\mathrm{x},\label{NMC-action}
\end{equation}
where $f_1(R)$ and $f_2(R)$ are arbitrary functions of the Ricci scalar $R$, $g$ is the determinant of the 4-dimensional spacetime metric $g_{\mu\nu}$, $\mathcal{L}$ is the matter Lagrangian density and $\kappa=\frac{1}{16\pi G}$, with $G$ being the Newton constant \cite{Bertolami:2007}. 

Varying this action with respect to the metric, the field equations take the form: 
\begin{equation}
    R_{\mu\nu}\Theta(R)-\left(\nabla_\mu \nabla_\nu -g_{\mu\nu}\Box\right)\Theta(R)-\frac{1}{2}g_{\mu\nu}f_1(R)=\frac{1}{2\kappa}f_2(R)T_{\mu\nu}, \label{NMC-field-eqs}
\end{equation}
where $\Theta(R)=F_1(R)+\left(F_2(R)/\kappa\right)\mathcal{L}$, such that $F_i(R)=\frac{\mathrm{d} f_i(R)}{\mathrm{d} R}$, with $i\in\{1,2\}$, $\Box=\nabla_\mu \nabla^\mu$ and $T_{\mu\nu}$ is the usual energy--momentum tensor build from the matter Lagrangian density. General Relativity is recovered for $f_1(R)=R$ and $f_2(R)=1$ as equations (\ref{NMC-field-eqs}) reduce then to Einstein's field equations. 

Applying the contracted covariant derivative in equations (\ref{NMC-field-eqs}), and using the contracted Bianchi identities, it is possible to obtain the relation
\begin{equation}
    \nabla_\mu T^{\mu\nu}=\frac{F_2(R)}{f_2(R)}\left( \mathcal{L}g^{\mu\nu}-T^{\mu\nu}\right)\nabla_\mu R. \label{NMC-Non-Cons-Eq}
\end{equation}

In this model, the energy--momentum tensor is, in general, not covariantly conserved. This feature has several implications, one of which is the lifting of the degeneracy in the choice of the matter Lagrangian density that gives rise to a perfect fluid energy--momentum tensor\cite{Bertolami:2008perfectfluids}. Several other implications of this theory have also been investigated. For instance, a generalization of the coupling can result in an extra force in the geodesic equation, which may mimic the effects of dark matter on galactic dynamics\cite{Bertolami:2007}. Additionally, general black hole solutions in this theory, that satisfy the standard energy conditions, have been derived and analyzed\cite{Bertolami:2015bhs}. At the cosmological level, this theory has also been explored. One example is the investigation of inflationary solutions driven by a scalar field in the presence of a non-minimal coupling between matter and curvature\cite{Gomes:2017}. Another study considers inflationary dynamics, where conformal symmetry is explicitly broken through the non-minimal coupling between curvature and the electromagnetic field, preventing the dilution of primordial magnetic fields throughout the inflationary epoch\cite{Bertolami:2022magnetic}.

So far, our discussion has been limited to connections that are both metric-compatible and free of torsion, as in General Relativity. However, this framework can be generalized, considering non-metricity properties. 

\section{Non-Minimally Coupled Weyl Connection Theory}
\label{NMCWCT-section}

Non-metricity describes a property of spacetime where the metric tensor does not remain covariantly constant, unlike in Riemannian geometry, where the covariant derivative of the metric is zero by definition. A particular case is the so-called Weyl connection gravity, where a non-vanishing vector field influences the geometry, being related to the metric through a specific coupling that modifies the connection, leading to non-metricity of the form: 
\begin{equation}
    D_\lambda g_{\mu\nu}=A_\lambda g_{\mu\nu},\label{covariant-deriv-eq}
\end{equation}
where $A_\lambda$ is the Weyl vector appears \cite{Weyl:1918ib} and the generalized covariant derivative is given by
\begin{equation}
	D_\lambda g_{\mu\nu}=\nabla_\lambda g_{\mu\nu}-\bar{\bar{\Gamma}}^\rho_{\mu\lambda}g_{\rho\nu}-\bar{\bar{\Gamma}}^\rho_{\nu\lambda}g_{\rho\mu},
 \label{relation1}
\end{equation}
where $\nabla_\lambda$ is the usual covariant derivative with Levi--Civita connection, and $\bar{\bar{\Gamma}}^\rho_{\mu\nu}=-\frac{1}{2}\delta^\rho_{\mu}A_\nu-\frac{1}{2}\delta^\rho_\nu A_\mu+\frac{1}{2}g_{\mu\nu}A^{\rho}$ is the deflection tensor which reflects the Weyl non-metricity. Taking into account the generalized connection, $\bar{\Gamma}^\rho_{\mu\nu}=\Gamma^\rho_{\mu\nu}+\bar{\bar{\Gamma}}^\rho_{\mu\nu}$, it is possible to deduce the generalized Ricci tensor, given by
\begin{equation}
\hspace{-1cm}	\bar{R}_{\mu\nu}=R_{\mu\nu}+\frac{1}{2}A_\mu A_\nu +\frac{1}{2}g_{\mu\nu} \left(\nabla_\lambda-A_\lambda\right)A^\lambda +\tilde{F}_{\mu\nu}+\frac{1}{2}\left(\nabla_\mu A_\nu+\nabla_\nu A_\mu\right)=R_{\mu\nu}+\bar{\bar{R}}_{\mu\nu},
	\label{Riccitensor}
\end{equation}
where $R_{\mu\nu}$ is the usual Ricci tensor and $\tilde{F}_{\mu\nu}=\partial_\mu A_\nu-\partial_\nu A_\mu=\nabla_\mu A_\nu -\nabla_\nu A_\mu$ is the Weyl field strength tensor. It is straightforward to observe that the trace of the generalized Ricci tensor, the scalar curvature associated with the Weyl connection, is given by
\begin{equation}
	\bar{R}=R+3\nabla_\lambda A^\lambda-\frac{3}{2}A_\lambda A^\lambda=R+\bar{\bar{R}},
    \label{RicciScalar}
\end{equation}
where $R$ is the usual Ricci~curvature.

Building upon these quantities, one may formulate a generalized version of the action (\ref{NMC-action}) by allowing $f_1(\bar{R})$ and $f_2(\bar{R})$ to explicitly depend on the generalized scalar curvature $\bar{R}$ \cite{Gomes:2019weyl}. Varying the action with respect to the vector field, up to boundary terms, yields an additional constraint-like equation given by
\begin{equation}
 	\nabla_\lambda \Theta(\bar{R})=-A_\lambda \Theta(\bar{R}). 
 	\label{eqconstraint}
 \end{equation}

Subsequently, varying the action with respect to the metric, it leads to the corresponding field equations: 
\begin{equation}
 	\left( R_{\mu\nu}+\bar{\bar{R}}_{(\mu\nu)} \right) \Theta(\bar{R})-\frac{1}{2}g_{\mu\nu}f_1(\bar{R})=\frac{f_2(\bar{R})}{2}T_{\mu\nu}.
 	\label{NMCWG-field-eqs}
 \end{equation}

A comparison between the field equations of the non-minimal coupling theory (\ref{NMC-field-eqs}) and those of the present theory (\ref{NMCWG-field-eqs}) reveals that the equations now take a second-order form, as a consequence of the constraint equation, in contrast to the usual fourth-order formulation.\\

In the same way as in equations (\ref{NMC-Non-Cons-Eq}), it is possible to derive a non-conservation law for energy--momentum tensor, given by 
\begin{adjustwidth}{-2.5cm}{0cm}
\begin{equation}
	\nabla_\mu T^{\mu\nu}=\frac{2}{f_2(\bar{R})}\left[ \frac{F_2(\bar{R})}{2}\left( g^{\mu\nu}\mathcal{L}-T^{\mu\nu} \right)\nabla_\mu R+\nabla_\mu(\Theta(\bar{R}) B^{\mu\nu})-\frac{1}{2}\left(F_1(\bar{R})g^{\mu\nu}+F_2(\bar{R})T^{\mu\nu}\right)\nabla_\mu \bar{\bar{R}} \right],
	\label{nonconserveq}
\end{equation}
\end{adjustwidth}
where $B^{\mu\nu}=\frac{3}{2}A^\mu A^\nu+\frac{3}{2}g^{\mu\nu}(\nabla_\lambda-A_\lambda)A^{\lambda}$. In contrast to previous model where the non-minimal curvature--matter coupling solely accounted for the energy--momentum non-conservation, the current framework reveals that non-metricity also plays a significant role, introducing further contributions to the exchange between geometry and matter sectors.

\subsection{Black Hole Solutions}\label{sec:blackholes}

A comprehensive assessment of any modified theory of gravity requires its application to a variety of scenarios. Studying solutions in these regimes enables an examination how key features of the theory, such as non-metricity and non-minimal curvature–matter couplings, manifest under strong-field conditions, where deviations from General Relativity may become significant. This chapter presents black hole solutions within the framework of the theory, based on the work by \cite{Lima:2024}, with a particular focus on the Schwarzschild-like and Reissner--Nordstrøm-like solutions.

\subsubsection{Schwarzschild-Like Black Hole}

In this section, we review two distinct vacuum solutions describing Schwarzschild-like black holes. Given the structure of the problem, only two types of Weyl vectors are allowed, namely: $A_\mu=\left( 0,A(r),0,0 \right)$, with $A(r)>0$, and $A_\mu=\left( A_0(r),A_1(r),0,0 \right)$, where $A_0(r)+A_1(r)>0$, and the functions $A_0(r)$ and $A_1(r)$ are related through the metric. The static line element in spherical coordinates is given by $ds^2=-g(r)dt^2+h(r)dr^2+r^2\left( d\theta^2+\sin^2(\theta)d\phi^2\right)$, where $g(r)$ and $h(r)$ are arbitrary functions of the distance, $r$.   

In vacuum, the model reduces to $f_1(\bar{R})=\alpha \bar{R}^2$, with $\alpha$ a constant. This form immediately implies that $\bar{R}=0$ and $\bar{R}_{\mu\nu}=0$, for all $\mu,\nu \in {0,1,2,3}$.

This condition leads to two possible solutions. In the first case, where the Weyl vector is taken as $A_\mu=\left( 0,A(r),0,0 \right)$, the solution is given by: 
\begin{subequations}\label{solution-radialSchw}
\begin{align}
	g(r)&=1-\frac{2M}{r}+\frac{2(\omega+3M)}{\omega}\frac{r}{\omega}+\frac{\omega+4M}{\omega}\left(\frac{r}{\omega}\right)^2,\\
    h(r)&=\frac{1+6\left( \frac{M}{\omega} \right)}{1-\frac{2M}{r}+\frac{2(\omega+3M)}{\omega}\frac{r}{\omega}+\frac{\omega+4M}{\omega}\left(\frac{r}{\omega}\right)^2},\\
    A(r)&=\frac{2}{r+\omega},
 \end{align}
\end{subequations}
where $\omega>0$ is the Weyl constant and $M>0$ is the black hole mass\cite{Lima:2024}.

Examining the solution, it is possible to see that close to the black hole, the metric component behaves as $g(r)\approx 1-\frac{2M}{r}$, which coincides with the well-known Schwarzschild solution in general relativity. On the other hand, at large distances, the component becomes $g(r)\approx 1+\frac{2(\omega+3M)}{\omega}\frac{r}{\omega}+\frac{\omega+4M}{\omega}\left(\frac{r}{\omega}\right)^2$, with an expansive behavior. This black hole solution possesses a single event horizon, located at $r_{H}=2M\frac{\omega}{4M+\omega}$. By calculating the Kretschmann invariant, which is the generalized Riemann squared tensor $K=\bar{R}_{\mu\nu\sigma\rho}\bar{R}^{\mu\nu\sigma\rho}$, it was observed that it diverges only at $r=0$ and the singularity is an essential one.\\

In the second case, where the Weyl vector is assumed to be $A_\mu=\left( A_0(r),A_1(r),0,0 \right)$, the solution reads as follows:
\begin{subequations}\label{SotulitonCase3.2}
\begin{align}
	g(r)&=1-\frac{2M}{r}+\frac{M}{2\omega}\left( \frac{r}{\omega} \right)-\frac{1}{4}\left( \frac{r}{\omega} \right)^2,\\
    h(r)&=\frac{1}{1-\frac{2M}{r}+\frac{M}{2\omega}\left( \frac{r}{\omega} \right)-\frac{1}{4}\left( \frac{r}{\omega} \right)^2},\\
    A_0(r)&=\frac{1}{\omega}\left( 1-\frac{2M}{r}  \right),\\
    A_1(r)&=\frac{2r}{r^2-4\omega^2},
 \end{align}
\end{subequations}
where $M>0$ is the black hole mass and $\omega>0$ is the Weyl constant\cite{Lima:2024}. 

This new solution has two possible event horizons: $r^{^{(M)}}_{H}=2M$ and $r^{^{(\omega)}}_{H}=2 \omega$. Using the Kretschmann invariant to assess the nature of the singularities in this result, it was found that the invariant diverges only at $r=0$. Therefore, $r=0$ corresponds to an essential singularity in the center of the black hole.\\

Having discussed the vacuum solutions, we now turn to the case where the matter Lagrangian density is non-vanishing. To capture the first non-trivial modifications to the geometry, a scenario was considered in which the matter content is modeled by a cosmological constant. For that, the Lagrangian density is defined by $\mathcal{L}^{^{(\Lambda)}}$. The corresponding field equations (\ref{NMCWG-field-eqs}) take the form: 
\begin{equation}
    \bar{R}_{\mu\nu} \left( F_1(\bar{R})-2\Lambda F_2(\bar{R})\right)-\frac{1}{2} g_{\mu\nu} \left(f_1(\bar{R})-2\Lambda f_2(\bar{R}) \right)=0. \label{SBH-CosmologicalConstant}
\end{equation}

By taking the trace of the field equations, the model is automatically defined by the relation $f_1(\bar{R})-2\Lambda f_2(\bar{R})=\alpha \bar{R}^2$, with $\alpha$ a constant. This, in turn, implies that $\bar{R}=0$ and $\bar{R}_{\mu\nu}=0$, for all $\mu,\nu \in {0,1,2,3}$. Therefore, the vacuum solutions, previously explained, remain valid even in the presence of a cosmological constant matter. One may look at this problem as a mathematical reparametrization of the model, considering $f(\bar{R})=f_1(\bar{R})-2\Lambda  f_2(\bar{R})$, as shown in Equation (\ref{SBH-CosmologicalConstant}). However, despite this equivalence, two possibilities are physically distinct: the cosmological constant term simply as a coefficient in the reparametrization or being associated with the vacuum energy. 

\subsubsection{Reissner--Nordstrøm-Like Black Hole}

In this section, we review one solution describing Reissner--Nordstrøm black holes, i.e., black holes that are static, and have mass and electric charge. For that, it is necessary to introduce the modified Maxwell equations, given by

\begin{equation}
    \nabla_\mu (f_2(\bar{R})F^{\mu\nu})=0, 
    \label{Maxwell_eqs}
\end{equation}
where $F_{\mu\nu}=\partial_\mu \Phi_\nu-\partial_\nu \Phi_\mu$ is the Faraday tensor and $\Phi_\mu$ is the electromagnetic four-potential\cite{Lima:2024}. In order to describe the gravitational field outside a charged, non-rotating, spherically symmetric body, the electrostatic four-potential $\Phi_\mu=(-\phi(r),0,0,0)$ was assumed, where $\phi(r)$ is the scalar potential. In this case, given the structure of the problem, only one form of Weyl vector is allowed: $A_\mu = (0, A(r), 0, 0)$, with $A(r)>0$.\\

The first consequence highlighted in Ref. \cite{Lima:2024} is that Reissner--Nordstrøm solution cannot be obtained in vacuum. However, if a cosmological constant background is considered, the solution to the problem is given by:

\begin{subequations}\label{solutionRN}
\begin{align}
        g(r)&=1-\frac{2M}{r}+\frac{\tilde{Q}^2}{r^2}+\frac{2\left(\omega^2+3M\omega+2\tilde{Q}^2\right)}{\omega^2} \frac{r}{\omega}+\frac{4M+\omega}{\omega}\left( \frac{r}{\omega} \right)^2,\\
        f(r)&=\frac{1+6\left(\frac{M}{\omega}+\frac{\tilde{Q}^2}{\omega^2}\right)}{1-\frac{2M}{r}+\frac{\tilde{Q}^2}{r^2}+\frac{2\left(\omega^2+3M\omega+2\tilde{Q}^2\right)}{\omega^2} \frac{r}{\omega}+\frac{4M+\omega}{\omega}\left( \frac{r}{\omega} \right)^2},\\
         A(r)&=\frac{2}{r+\omega},\\
        \phi(r)&=\frac{\tilde{Q}}{r},
\end{align}
\end{subequations}
where $\omega>0$ is the Weyl constant, M is the mass of the black hole, $Q$ is the usual charge, and $\tilde{Q}$ is a dressed charge, related by the expression $Q^2=\zeta^2 \tilde{Q}^2 \left( 1+6\left(\frac{M}{\omega}+\frac{\tilde{Q}^2}{\omega^2}\right) \right)^{-1}$. The model was defined by $f_1(\bar{R})=\alpha \bar{R}^2+2\Lambda \zeta$ and $f_2(\bar{R})=\zeta$, where $\alpha$ and $\zeta$ are related constants. 

Examining the solution, one can see that in the limit $\tilde{Q}\rightarrow0$, this Reissner--Nordstrøm solution converge to the previous solution (\ref{solution-radialSchw}). Depending on the values chosen for the parameters, this solution may exhibit two, one, or no event horizons. Using the Kretschmann invariant to assess the nature of the singularities in this case, it was shown that the invariant diverges only at $r = 0$. Therefore, $r = 0$ corresponds to an essential singularity in the center of the black hole.

\section{Modified Friedmann Equations in the Minimal Case}\label{sec:modifiedFriedmann}

In this work, our goal is to investigate the impact of Weyl non-metricity in the case of scalar perturbations. For this purpose, we restrict the analysis to the minimal case, $f_2(\bar{R})=1$. Accordingly, we set $\Theta(\bar{R})=F_1(\bar{R})\equiv F(\bar{R})$. For the sake of simplicity, we drop the index $1$ in the free function. We shall also consider a cosmological line element given in conformal time:
\begin{equation}
    d\tilde{s}^2=a^2(\eta) \left( -d\eta^2+\delta_{ij}dx^i dx^j \right),
    \label{Background-Metric}
\end{equation}
where $a(\eta)$ is the scalar factor in comoving time. Moreover, the Weyl vector field should have the following cosmological principle compatible ansatz: $\tilde{A}_\mu=\left(\tilde{A}(\eta),0,0,0\right)$. In addition, we shall assume that the energy/matter content of the Universe is well described by a perfect fluid, whose energy-momentum tensor is of the form $T_{\mu\nu}=a^2(\eta) \text{diag}(\tilde{\rho},\tilde{P},\tilde{P},\tilde{P})$. So the field equations (\ref{NMCWG-field-eqs}) become:
\begin{subequations}
\begin{align}
\left( 3\tilde{A}'-6\mathcal{H}' \right) F(\bar{R}) + a^{2}(\eta)f(\bar{R}) &= a^{2}(\eta) \tilde{\rho}\\
\left( 2\mathcal{H}'-\tilde{A}'+4\mathcal{H}^2-4\mathcal{H}\tilde{A}+\tilde{A}^2 \right)F(\bar{R}) + a^{2}(\eta)f(\bar{R}) &= a^{2}(\eta) \tilde{P}
\label{Background-field-equations}
\end{align}
\end{subequations}

Considering this background, the only non-vanishing equation in the non-conservation law for the energy--momentum tensor (\ref{nonconserveq}) is
\begin{equation}
    \tilde{\rho}'+3\mathcal{H}(\tilde{\rho}+\tilde{P})=0.
\end{equation}

For this background, using the metric (\ref{Background-Metric}) and the non-perturbed Weyl vector, and considering a minimal coupling, the so-called non-conservation law for the energy–-momentum tensor is, in fact, a conservation law since the right-hand side of equation (\ref{nonconserveq}) is zero.

We are now able to perturb the cosmological equations, as well as the non-conservation equations for the energy–-momentum tensor.

\section{Cosmological Perturbations}\label{sec:cosmologicalperturbations}

Scalar cosmological perturbations follows from the following decomposition of the metric field up to first order, and resorting to the so-called gauge invariant Bardeen scalars, $\Phi$ and $\Psi$:
\begin{equation}
    	ds^2= a^2(\eta) \left( -(1+2\Psi)d\eta^2 +(1-2\Phi)\delta_{ij}dx^idx^j \right),
\end{equation}
together with the decomposition of the Weyl vector field ansatz:
\begin{equation}
    A_\mu=\left(\tilde{A}(\eta)+\delta A_0(\eta,\mathrm{x}),\delta A_1(\eta,\mathrm{x}),\delta A_2(\eta,\mathrm{x}),\delta A_3(\eta,\mathrm{x})\right),
\end{equation}
from which one can compute all the curvature associated quantities up to first order, which we present in the \ref{sec:annex}.\\

Starting from constraint (\ref{eqconstraint}), in the non-perturbative regime we have the relation $\partial_\lambda \Theta(\tilde{\bar{R}})=-\tilde{A}_\lambda \Theta(\tilde{\bar{R}})$. But now, for the perturbative part the constraint takes the form
\begin{equation}
    \partial_\lambda (\delta\bar{\Theta}) = -\delta A_\lambda \Theta(\tilde{\bar{R}})-\tilde{A}_\lambda\delta \bar{\Theta},
    \label{perturbed-constraint}
\end{equation}
where $\delta \bar{\Theta}$ is such that $\Theta(\bar{R})=\Theta(\tilde{\bar{R}})+\delta\bar{\Theta}$. Thus, the 0 equation gives us the relation
\begin{equation}
\delta\bar{\Theta}' = -\delta A_0 \tilde{\bar{\Theta}}-\tilde{A}\delta \bar{\Theta},
\end{equation}
and the $i$ equations are such that
\begin{equation}
    \partial_i(\delta\bar{\Theta})=-\delta A_i\tilde{\bar{\Theta}}.
\end{equation}

In order to analyze the behavior of the scalar, vector and tensorial modes in this configuration, it is convenient to use the SVT decomposition. So, we consider that $\delta A_i=\partial_i\Sigma+\delta \hat{A}_i$, where $\partial^k \hat{A}_k=0$. 

Applying this decomposition in the previous $i$--constraint, we can see that the vector component should be zero ($ \hat{A}_i=0$) and the relation takes the form 
\begin{equation}
    \delta \bar{\Theta}=-\Sigma \hspace{0.1cm} \Theta(\tilde{\bar{R}}).
    \label{restriction-i}
\end{equation}

Applying this new result in the $0$--constraint, we obtain the relation 
\begin{equation}
    \Sigma'=\delta A_0. \\
\end{equation}

We can now obtain the perturbed field equations also performing the SVT decomposition of the energy–momentum tensor. Combining all the information discussed so far, the perturbed field equations, in the Fourier space, can be written as
\begin{subequations}
\begin{gather}
\hspace{-2cm} \Bigg(2\kappa^2(2\Phi+\Sigma)+6(2\mathcal{H}-\tilde{A})\Phi'-3(2\mathcal{H}'-\tilde{A}')\Sigma+3(4\mathcal{H}^2-\tilde{A}(4\mathcal{H}-\tilde{A}))\Psi+3(2\mathcal{H}-\tilde{A})\delta A_0 \Bigg)F(\bar{R})=-a^2(\eta)\delta\rho,\\
\left( 4\Phi'+2\delta A_0+(\tilde{A}-2\mathcal{H})(\Sigma-2\Psi) \right)F(\bar{R}) = -a^2(\eta)(\tilde{\rho}+\tilde{P})v, \\
2(\Phi-\Psi+\Sigma)F(\bar{R})=a^2(\eta)\Pi,\\
\hspace{-2cm} \Bigg(\frac{4}{3}\kappa^2(\Phi-\Psi+\Sigma)+4\Phi''+2(2\mathcal{H}-\tilde{A})(2\Phi'+\Psi')+2\delta A_0'+\left(2(\mathcal{H}'+2\mathcal{H}^2)-\tilde{A}'-\tilde{A}(4\mathcal{H}-\tilde{A})\right)(\Psi-\Sigma)+\nonumber\\
+3(2\mathcal{H}'-\tilde{A}')\Psi+(2\mathcal{H}-\tilde{A})\delta A_0 \Bigg)F(\bar{R})=a^2(\eta)\delta P,
\end{gather}
\end{subequations}
where $v$ and $\Pi$ are such that $v_i=\partial_iv$ and $\Pi_{ij}=\partial_i\partial_j\Pi-\frac{1}{3}\delta_{ij}\partial^2\Pi$. Due to the chosen metric and the absence of Weyl vector–induced vector modes, the tensorial and vector modes of the energy–-momentum tensor vanish. It is straightforward to note that $\delta A_0$ has the same dimension as $\tilde{A}$ and $\mathcal{H}$, while $\Sigma$ has the same dimension as the scalar metric perturbations $\Phi$ and $\Psi$.\\ 

Considering the conservation equation of the energy–momentum tensor (\ref{NMC-Non-Cons-Eq}), and performing the necessary calculations, one finds the following perturbed equations, in the Fourier space:

\begin{subequations}
\begin{gather}
\hspace{-1cm} \delta\rho'-3(\tilde{\rho}+\tilde{P})\Phi'- \kappa^2 (\tilde{\rho}+\tilde{P})v+3\mathcal{H}(\delta\rho+\delta P) = 0\\
\hspace{-1cm} (\tilde{\rho}+\tilde{P})v'+\left(\tilde{P}'+\mathcal{H}(\tilde{\rho}+\tilde{P}) \right)v+\delta P+(\tilde{\rho}+\tilde{P})\Psi-\frac{2}{3}\kappa^2\Pi = \nonumber\\
=6 a^{-2}(\eta) F(\bar{R}) \Big( \tilde{A}\delta A_0+\left(\tilde{A}'+\tilde{A}(2\mathcal{H}-\tilde{A})\right)\Sigma \Big).
\end{gather}
\end{subequations}

\section{Conclusions} \label{sec:conclusions}

We have reviewed some properties of the non-minimally coupled Weyl gravity model, including black hole solutions. This model presents Schwarzschild-like solutions both in vacuum and in the presence of a cosmological constant-like term. However, Reissner--Nordstrøm-like solutions only exist in the presence of matter. All the singularities are essential as computed from the Kretschmann invariant.

We have analyzed in the minimal case, namely $f_2(\bar{R})=1$, the cosmological field equations which generalize the Friedmann equations. These have extra terms associated with the Weyl vector field. However, for the choices made for the ansatz of the Weyl vector and the minimal case, there is a conservation law for the energy density in a cosmological background as in GR.

In what concerns scalar cosmological perturbations, i.e., around a Friedmann-Robertson-Walker metric, the results are highly non-trivial, even in the minimal case, as there are new terms associated with the perturbation of the Weyl vector. The perturbed equations from the non-conservation of the energy--momentum tensor are also non-trivial, offering a new rich environment to be further scrutinized.

The scalar perturbations found in this work will be essential in future work when compared with data, and when deriving the generalized virial theorem, also known as Layzer-Irvine equation, a powerful cosmological tool for numerical simulations and for testing dark matter \cite{Layzer,Irvine,Winther:2013,Gomes:2013,Gomes:2014,Gomes:2025xwg}.

\appendix

\section{Geometric Quantities}\label{sec:annex}
In this section, we present several geometric quantities written up to linear order in perturbation theory. We can thus check the influence of the non-metricity property in those. For that, let's start to define the perturbed metric like:
\begin{equation}
    	ds^2= a^2(\eta) \left( -(1+2\Psi)d\eta^2 +(1-2\Phi)\delta_{ij}dx^idx^j \right),
\end{equation}
where $\Psi$ and $\Phi$ are scalar perturbations. We shall denote all quantities computed at the background with a tilde, so $g_{\mu\nu}=\tilde{g}_{\mu\nu}+\delta g_{\mu\nu}$.\\

Considering the non-perturbed field equations and the symmetry of the non-perturbed metric, a possible non-perturbed Weyl vector, given the constraint (\ref{eqconstraint}), is $\tilde{A}_\mu=\left(\tilde{A}(\eta),0,0,0\right)$. Thus, the perturbed Weyl vector is given by:

\begin{equation}
    A_\mu=\left(\tilde{A}(\eta)+\delta A_0(\eta,\mathrm{x}),\delta A_1(\eta,\mathrm{x}),\delta A_2(\eta,\mathrm{x}),\delta A_3(\eta,\mathrm{x})\right).
\end{equation}
To simplify the notation, let us write $\delta A_i(\eta,\mathrm{x})$ only as $\delta A_i$, for all values of $i$ and $\tilde{A}(\eta)$ only as $\tilde{A}$. 

Given this configuration, the non-vanishing corrections in Christoffel symbols are given by
\begin{subequations}
\begin{align}
\bar{\bar{\Gamma}}^0_{00}&=\tilde{\bar{\bar{\Gamma}}}^0_{00}+\delta\bar{\bar{\Gamma}}^0_{00}=-\frac{1}{2}\tilde{A}-\frac{1}{2}\delta A_0,\\
\bar{\bar{\Gamma}}^0_{0i}&=\delta\bar{\bar{\Gamma}}^0_{0i}=-\frac{1}{2}\delta A_i,\\
\bar{\bar{\Gamma}}^0_{ij}&=\tilde{\bar{\bar{\Gamma}}}^0_{ij}+\delta\bar{\bar{\Gamma}}^0_{ij}=-\frac{1}{2}\tilde{A} \delta_{ij}+\left(-\frac{1}{2}\delta A_0+(\Psi+\Phi)\tilde{A} \right)\delta_{ij},\\
\bar{\bar{\Gamma}}^i_{00}&=\delta\bar{\bar{\Gamma}}^i_{00}=-\frac{1}{2}\delta A_j \delta^{ji},\\
\bar{\bar{\Gamma}}^i_{0j}&=\tilde{\bar{\bar{\Gamma}}}^i_{0j}+\delta\bar{\bar{\Gamma}}^i_{0j}=-\frac{1}{2}\tilde{A}\delta^i_j-\frac{1}{2}\delta A_0 \delta^i_j,\\
\bar{\bar{\Gamma}}^i_{jk}&=\delta\bar{\bar{\Gamma}}^i_{jk}=-\frac{1}{2}\delta^i_j\delta A_k -\frac{1}{2}\delta^i_k \delta A_j+\frac{1}{2}\delta_{jk}\delta^{il} \delta A_l. 
\end{align}
\end{subequations}

Considering (\ref{Riccitensor}), the non-vanishing correction Ricci tensor components take the form
\begin{subequations}
\begin{align}
	\bar{\bar{R}}_{00}&=\tilde{\bar{\bar{R}}}_{00}+\delta \bar{\bar{R}}_{00}=\frac{3}{2}\tilde{A}'+\frac{3}{2}\delta A_0'-\frac{3}{2}(\Psi'+\Phi')\tilde{A}-\frac{1}{2}\partial^i(\delta A_i),\\
	\bar{\bar{R}}_{(0i)}&=\delta \bar{\bar{R}}_{(0i)}=\frac{1}{2}\delta A_i'+\frac{1}{2}\partial_i(\delta A_0)+\frac{1}{2}(\tilde{A}-2\mathcal{H})\delta A_i-\partial_i\Psi \tilde{A},\\
	\bar{\bar{R}}_{(ij)}&=\tilde{\bar{\bar{R}}}_{(ij)}+\delta \bar{\bar{R}}_{(ij)}=\Big( -\frac{1}{2}\tilde{A}'-2\mathcal{H}\tilde{A}+\frac{1}{2}\tilde{A}^2 \Big)\delta_{ij} + \Big( -\frac{1}{2}\delta A_0'+\frac{1}{2}(\Psi'+5\Phi')\tilde{A}+\nonumber\\ 
	&\hspace{-0.5cm} +\frac{1}{2}\partial^i(\delta A_i)+(\tilde{A}-2\mathcal{H})\delta A_0+(\Psi+\Phi)(\tilde{A}'+\tilde{A}(4\mathcal{H}-\tilde{A})) \Big)\delta_{ij}+\frac{1}{2}\big( \partial_i(\delta A_j)+\partial_j(\delta A_i) \big). 
\end{align}
\end{subequations}

Now, considering the relation (\ref{RicciScalar}), the Ricci Scalar correction is given by
\begin{subequations}
\begin{align}
    \bar{\bar{R}}&=\tilde{\bar{\bar{R}}}+\delta \bar{\bar{R}}= a^{-2}(\eta)\Big( -3\tilde{A}'+\frac{3}{2}\tilde{A}(\tilde{A}-4\mathcal{H}) + 3 \big( -\delta A_0'+(\Psi'+3\Phi')\tilde{A}+\nonumber\\
    &\hspace{-0.5cm} +\partial^k(\delta A_k)+\delta A_0(\tilde{A}-2\mathcal{H})+\Psi(2\tilde{A}'+ \tilde{A}(4\mathcal{H}-\tilde{A}))  \big)\Big).
\end{align}
\end{subequations}

In terms of the matter content, we will assume that, in a non-perturbative regime, the energy--momentum tensor is like a perfect fluid. So, in the perturbative regime the energy--momentum tensor components are given by

\begin{subequations}
\begin{align}
	T_{00}&=\tilde{T}_{00}+\delta T_{00}=a^2(\eta) \tilde{\rho}+a^2(\eta) \left( \delta \rho+2\Psi \tilde{\rho} \right),\\
	T_{0i}&=\delta T_{0i}=-a^2(\eta)(\tilde{\rho}+\tilde{P})v_i,\\
	T_{ij}&=\tilde{T}_{ij}+\delta T_{ij}=a^2(\eta)\tilde{P}\delta_{ij}+a^2(\eta)\left( (\delta P-2\Phi \tilde{P})\delta_{ij} +\Pi_{ij} \right), 
\end{align}
\end{subequations}
where $\Pi^i_i=0$.\\

\section*{Acknowledgments}

M.L. acknowledges support from Fundo Regional da Ciência e Tecnologia and Azores Government through the Fellowship M3.1.a/F/031/2022, and from FCT/Portugal through CAMGSD, IST-ID, projects UIDB/04459/2020 and UIDP/04459/2020 and the H2020-MSCA-2022-SE project EinsteinWaves, GA No.101131233. C.G. acknowledges support from Fundação para a Ciência e a Tecnologia through UIDB/04650/2025 - Centro de Física das Universidades do Minho e do Porto.

\section*{ORCID}
\noindent M. Lima \url{https://orcid.org/0000-0003-2231-4978}

\noindent C. Gomes \url{https://orcid.org/0000-0001-6022-459X}

\bibliographystyle{ws-mpla}
\bibliography{sample}
\end{document}